\documentclass{article}
\usepackage{amsmath}
\usepackage{amssymb}
\usepackage{epsf}
\usepackage{graphicx}
\usepackage{setspace}
\usepackage{overcite}
\usepackage{caption}

\usepackage{times}

\date{}
\begin{document}

\title{Evidence for charge-vortex duality at the LaAlO$_3$/SrTiO$_3$ interface}

\author{M. M. Mehta$^1$, D.A. Dikin$^1$, C.W. Bark$^2$, S. Ryu$^2$, C.M. Folkman$^2$, C.B. Eom$^2$ \& V. Chandrasekhar$^1$}

\footnotetext[1]{Department of Physics and Astronomy, Northwestern
University, Evanston, IL 60208, USA}

\footnotetext[2]{Department of Materials Science and Engineering
University of Wisconsin-Madison, Madison, WI 53706, USA}

\baselineskip24pt

\maketitle

\large
\textbf{
The concept of duality has proved extremely powerful in extending our understanding in many areas of physics\cite{girvin,fisher2}. 
Charge-vortex duality has been proposed \cite{fisher,fazio}
as a model to understand the superconductor to insulator transition (SIT) in disordered thin films\cite{haviland,hebard} and  Josephson junction arrays 
\cite{geerlings,vdzant,rimberg}. 
In this model, on the superconducting side, one has delocalized Cooper pairs but localized vortices; while on the insulating side, one has localized Cooper pairs but mobile vortices. Here we show a new experimental manifestation of this duality in 
the electron gas that forms at the interface between LaAlO$_3$ (LAO) and SrTiO$_3$ (STO)\cite{ohtomo,thiel,reyren,caviglia,dikin}. 
The effect is due to the motion of vortices generated by the magnetization dynamics of the ferromagnet that also forms at the 
same interface\cite{dikin,bert,li}, which results in an increase in resistance on the superconducting side of the transition, but an 
increase in conductance on the insulating side.}

\normalsize

The two dimensional SIT has been studied extensively \cite{geerlings,hebard, haviland} as experimentally it is one of the few easily 
accessible manifestations of a quantum phase transition. The two 
dimensional electron gas formed at the LAO-STO interface has been shown to undergo a SIT as the density of the carriers is tuned  
\cite{caviglia,dikin,schneider}. In addition to superconductivity, our earlier work showed evidence of ferromagnetism, which  manifests itself as a 
hysteresis in measurements of the magnetoresistance (MR), Hall effect and
superconducting phase boundary \cite{dikin}.  Two other recent studies have 
confirmed the coexistence of superconductivity and magnetism in similar samples \cite{bert,li}. The experimental evidence obtained so far 
\cite{dikin,bert} indicates that the effect of the ferromagnetism on the electrical transport properties of the 
system can be accounted for by considering only the magnetic field produced by the ferromagnet. As we show below, this close proximity of the 
ferromagnet and the superconductor allows for the observation of a novel demonstration of charge-vortex duality.

The samples in this work had 10 unit cells (uc) of LAO grown by 
pulsed laser deposition on TiO$_2$ terminated (001) STO single crystal substrates \cite{park,bark}.  Details of sample fabrication
were discussed in an earlier publication \cite{dikin} (see also Supplementary Information (SI)). The 
interaction between ferromagnetism and superconductivity can be seen in  
MR measurements in the superconducting state. Figures 1a and 1b show the MR in a magnetic field parallel and perpendicular to the
plane of the sample respectively,
at $T =$ 50 mK and $V_g =$ 80 V, a gate voltage at which the sample is superconducting. Consider first the case of a parallel magnetic 
field, $H_\parallel$.  As has been seen earlier for 
this sample in perpendicular magnetic fields \cite{dikin}, the parallel field MR is hysteretic due to the presence of ferromagnetic order 
(inset of Fig. 1a). 
Since $H_\parallel$ is in the plane of the 
two-dimensional superconductor, the critical field is large ($H_{c\parallel} \sim$ 1.2 T, see Fig. S1 of the
SI),  so that one expects the MR to be small in this field range.  In fact, the overall background MR is indeed quite small:  
most of the contribution to the MR is due to a set of twin peaks at $\pm$12 mT.
For clarity, we show in Fig. 1a the MR for only one sweep direction. As
$H_\parallel$ is increased from negative values, there is a small 
decrease in the resistance of the sample until $H_\parallel$=0.  The peak in the MR develops just as $H_\parallel$ becomes positive for this 
field direction, but then rapidly dies out as $H_\parallel$ is increased
further.  The most surprising fact about the MR of Fig. 1a is that the amplitude
of the peak depends on the 
sweep rate of $H_\parallel$: the amplitude \textit{increases} as the field
sweep rate is increased.  For very slow sweep rates, the peak is almost absent,
and the amplitude of the overall MR is very small, as expected from our
earlier discussion.

The rate dependence of the MR can be understood if we consider the influence of the domain wall of the ferromagnet on the superconductor, as shown
schematically in Fig. 2a.  In the simplest model, as $H_\parallel$ is 
swept from a large negative to a large positive value, the magnetization of the ferromagnetic layer reverses by forming a 
domain wall that propagates from one end of the sample to the other once $H_\parallel >$ 0.  If the domain wall is a 
Bloch wall \cite{kittel}, the magnetization points out of the plane of the interface, and the conducting layer sees a magnetic field perpendicular 
to the interface that is localized near the domain wall, and travels along with
it.  We emphasize that for the real sample, the magnetization dynamics are likely 
far more complicated, but the end result of a localized magnetic 
field perpendicular to the plane of the interface that moves as $H_\parallel$ is swept is the same.  At the field sweep rates in the experiment, 
the motion of the domains is quasi-static, a fact that can be confirmed by micromagnetic simulations.  Consequently, the time dependence of the 
localized perpendicular magnetic field is determined only by the sweep rate of the external magnetic field.  
The moving perpendicular component of the field in turn generates 
moving vortices in the superconductor. For our samples, which can be thought of as granular superconductors, moving vortices will result in a change 
in resistance\cite{jaeger} as the vortices cross the weak links between superconducting grains (see section 6, Supplementary Information).  This resistance change is 
proportional to the field sweep rate $\dot{B}$ and exponentially dependent on $E_J$ and hence on the critical current $I_c$.  One would therefore 
expect an increase in amplitude of the resistance peak with increasing field sweep rate at a specific gate voltage, and whose amplitude varies 
exponentially with $I_c$ at a specific sweep rate.  This is exactly what is seen in our experiments (lower part of Fig. 4a, and Fig. 4c).

Similar behaviour is also observed for the MR in a perpendicular external magnetic field $H_\perp$ (Fig. 1b). Given the two-dimensional 
nature of the ferromagnet \cite{lee}, it is unlikely that the perpendicular magnetic fields applied can cant the moments appreciably out of plane.  
However, a Bloch domain wall may form at the interface, and it is domains within this Bloch wall that are reoriented in an external magnetic field, 
as shown in Fig. 2b.  Such behavior has been seen before in thin magnetic films
\cite{kim}.  (We stress again that this picture is highly simplified, and the real reversal process is likely much more complicated.)  $H_\perp$ itself leads to a large overall MR, with a minimum at
zero field. 
The field from the ferromagnet gives rise to a sweep-rate dependent peak at 
$H_\perp \sim \pm$ 15 mT, and a second, less prominent peak at $H_\perp \sim \pm$ 48 mT, which is likely due to more complex magnetization 
dynamics in perpendicular field.  Further evidence that the hysteresis and the dips are due to magnetization dynamics in the ferromagnet can 
be seen from the MR in crossed parallel and perpendicular fields. When one measures the background MR in a perpendicular field in the presence 
of a large constant parallel field 
that aligns the magnetization of the ferromagnet, no hysteresis or rate
dependence is observed (Fig. S2 of the SI).

In the LAO/STO interface system, the superconducting transition can be tuned by
$V_g$ \cite{caviglia,dikin}. 
Early studies of LAO/STO interfaces identified this as a SIT \cite{caviglia}, which has been studied extensively in the past in thin superconducting 
films as a function of the film thickness or an applied magnetic field \cite{haviland,hebard}. However, in the LAO/STO system, it appears difficult to 
tune the system deep into the insulating regime with a gate voltage \cite{schneider}.  Nevertheless, measurements of the 
current dependent differential resistance indicate that the transition is to a weakly insulating state where regions of superconductivity get 
increasingly isolated from one another as $V_g$ is decreased (see Fig. S4 of the SI).  Thus, as we noted above, it makes sense to model our sample as a granular superconductor consisting of a  
network of superconducting islands whose phase is coupled through the Josephson effect, parametrized by the Josephson coupling 
energy $E_J$, and a Coulomb charging energy $E_c$, related to the energy cost of adding a Cooper pair to an island, \cite{fazio,geerlings}.   (For a random network of islands, these quantities are averages over the network.)
Changing $V_g$ in our sample tunes the transition by modifying the ratio between $E_J$ and $E_c$. The superconducting and insulating 
regimes are duals of each other, and this duality is predicted to manifest itself through an interchange between specific measurable 
quantities \cite{girvin,fisher2,fisher,fazio}. For example, the current $I$ and voltage $V$ are interchanged across the SIT, so that the 
$I$-$V$ curve in the superconducting phase is similar in shape to the $V$-$I$ curve in the insulating state. Such dual $I$-$V$ characteristics 
have been observed in Josephson junction arrays and disordered thin films \cite{geerlings,vdzant,rimberg}.

In our samples, as noted above, it appears difficult to tune the system deep into the insulating regime, so that observing this signature of 
charge-vortex duality is difficult \cite{reyren,dikin} (see the discussion in the SI). However, the interplay between ferromagnetism and  
superconductivity at the LAO/STO interface results in a unique manifestation of charge-vortex duality associated with moving ferromagnetic 
domain walls that has not been observed in other systems. 

Figures 1c and 1d demonstrate this new behaviour.  Figure 1c shows the equivalent of the data of Fig. 1a but at a gate voltage of $V_g =$ 
-100 V, placing the sample on the insulating side of the transition.  At the same magnetic fields at which a sweep-rate dependent peak was 
seen in Fig. 1a, a sweep-rate dependent \textit{dip} is observed.  As with the peak, the magnitude of the dip increases with increasing sweep 
rate.  Similar behavior is observed in perpendicular magnetic field (Fig. 1d). Note that in perpendicular field, one even observes two sets of 
dips (at $\pm$ 15 mT and $\pm$48 mT), mirroring the behavior seen in Fig. 1b. In the charge-vortex duality model, current and voltage are switched 
in going across the SIT, so that conductance and resistance are also switched.  Thus, where one observes peaks in resistance on the 
superconducting side, one should observe peaks in conductance (or dips in resistance) on the insulating side.  This is exactly what is observed.  
As $V_g$ is changed from +80 V to -100 V, the peaks in the MR change to dips (Figs. 3a and b).  A more striking graphical demonstration of the 
SIT can be seen if we plot the rate dependence of the resistance at the peaks or dips at different gate voltages normalized to their values at 
the highest sweep rate (Figs. 4a and b).  We note that the resistance per square, $R_S$, at which the transition occurs is approximately 
2.1 k$\Omega$, less than the universal resistance value of $h/4e^2\sim$6.45 k$\Omega$ expected for the SIT \cite{fisher2}.  

The peaks in the MR on the superconducting side are due to the magnetic field associated with a moving nonplanar magnetization
in the ferromagnet as the external magnetic field is swept.  Given the fact that the position of the peaks and dips coincide for both 
$H_\perp$ and $H_\parallel$, it is clear that the dips also have the same origin. How would this field give rise to an increase in 
conductance on the insulating side?  If the non-superconducting regime was simply a weakly localized metal, it would be hard to explain 
the dips in the MR that we observe (for a discussion see the SI). These dips can be explained if one models the insulating side of the 
transition as one where the Cooper pairs are localized on 
isolated superconducting islands, and a Cooper pair requires an energy $E_c$ to transfer between two neighboring islands, even though $E_c$ 
might be very small.  Classically, the moving nonplanar magnetization of the ferromagnet gives rise to a moving (i.e., time-dependent) 
field, which in turn gives rise to an electric field $\mathcal{E}_H = -\partial A/\partial t$ through Faraday's law, , where $A$ is the vector 
potential associated with the field.  The voltage difference $V=\mathcal{E}_H d$ generated between two
neighboring superconducting islands separated by a distance $d$ can exceed the charging energy $E_c$, $V>E_c/2e$, so that the conductance of 
the system is momentarily increased.  In contrast to the superconducting regime, this model predicts that the magnitude of the dip should depend exponentially on the sweep rate $\dot{B}$ (see Supplementary Information), a prediction that is borne out by the data (Fig. 4d). 

In summary, the LAO/STO interface, with its unusual combination of superconductivity and ferromagnetism, provides a unique signature of 
charge-vortex duality in the superconductor-insulator transition.  It would be interesting to see if similar signatures can be observed in artificially 
fabricated hybrid ferromagnet-superconductor structures.

\begin{figure}[h]
\begin{center}

\includegraphics[width=0.9\textwidth]{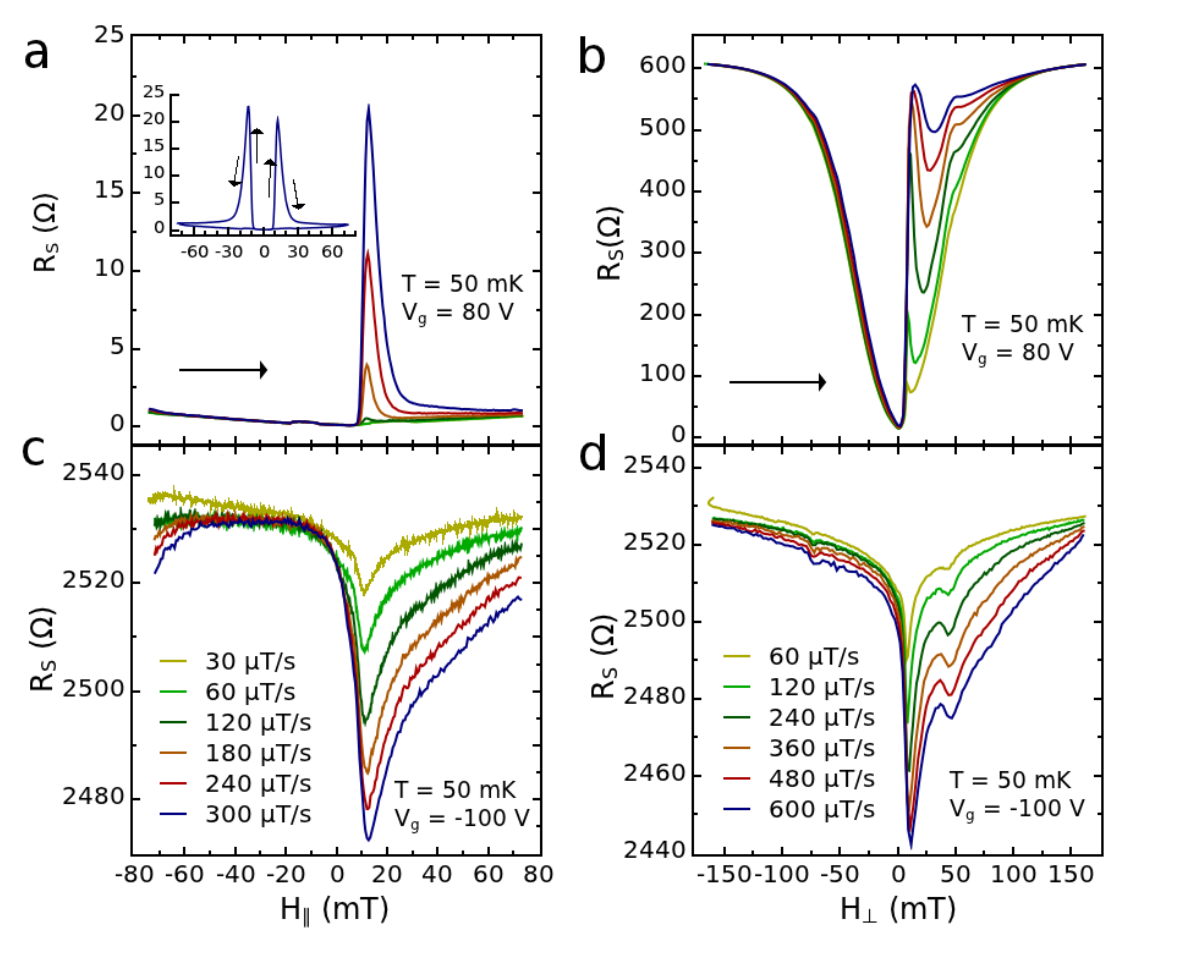}
\caption{\textbf{MR in the superconducting and insulating regimes.}
\textbf{a,} Parallel field MR as a function of different rates in the superconducting regime at $V_g =$ 80 V. Data for only one field sweep direction 
is shown for clarity. Arrow indicates the direction of field sweep. The inset shows the MR for forward and backward field sweep directions 
at a sweep rate of 300 $\mu$T/s. \textbf{b,} Perpendicular field MR for fields swept from 
negative to positive values for different field sweep rates for $V_g =$ 80 V.
\textbf{c,} Parallel field MR for $V_g =$ -100 V.
\textbf{d,} Perpendicular field MR for $V_g =$ -100 V. }
 
\end{center}
\label{fig1}
\end{figure}  

\begin{figure}[h]
\begin{center}
\includegraphics[width=0.9\textwidth]{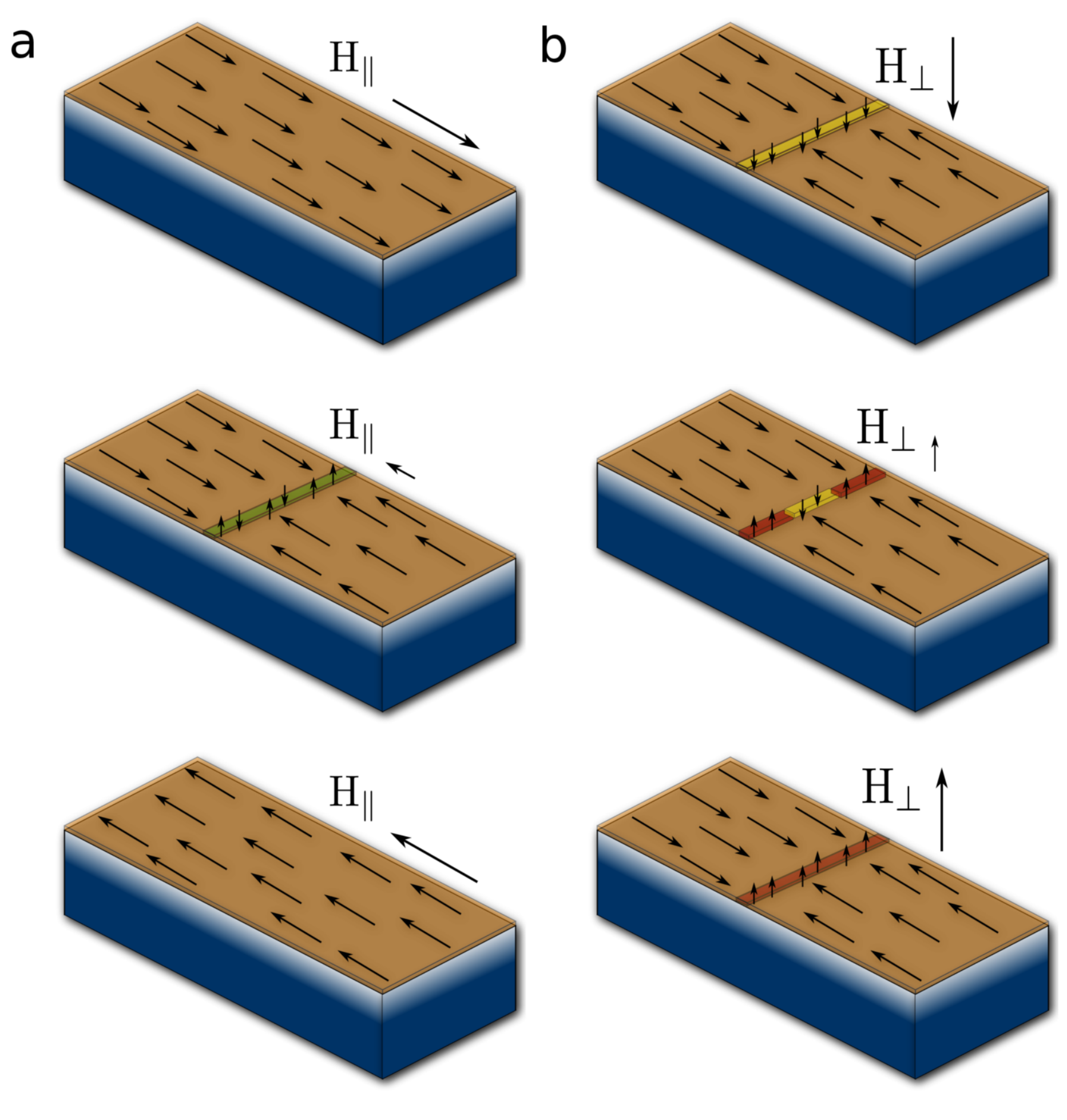}

%\vspace{-1cm}
\caption{\textbf{Magnetization dynamics of the ferromagnet.} \textbf{a,} Schematic of the system at different field values in parallel field. The 
top layer is the ferromagnet, the superconductor is shown through a gradient as extending some distance into the STO. 
The saturated configurations 
are shown at the two extremes when a high field is applied. Magnetization reversal by 
means of domain wall propagation in the ferromagnet for parallel field is depicted at a field where the peak in MR occurs. The domain wall 
induces vortices in the superconductor due to its perpendicular component of the field. It should be noted that the actual magnetization dynamics 
are more complicated in the real system, i.e. there are multiple domin walls moving in various directions. \textbf{b,} Schematic of the 
magnetization state of the system at different perpendicular fields.  Due to the shape anisotropy of the system, the majority of the moments lie in plane, but the external magnetic field orients the direction of the perpendicular component of the magnetization of the domain wall.} 

\end{center}
\label{fig2}
\end{figure} 

\begin{figure}[h]
\begin{center}
\includegraphics[width=0.9\textwidth]{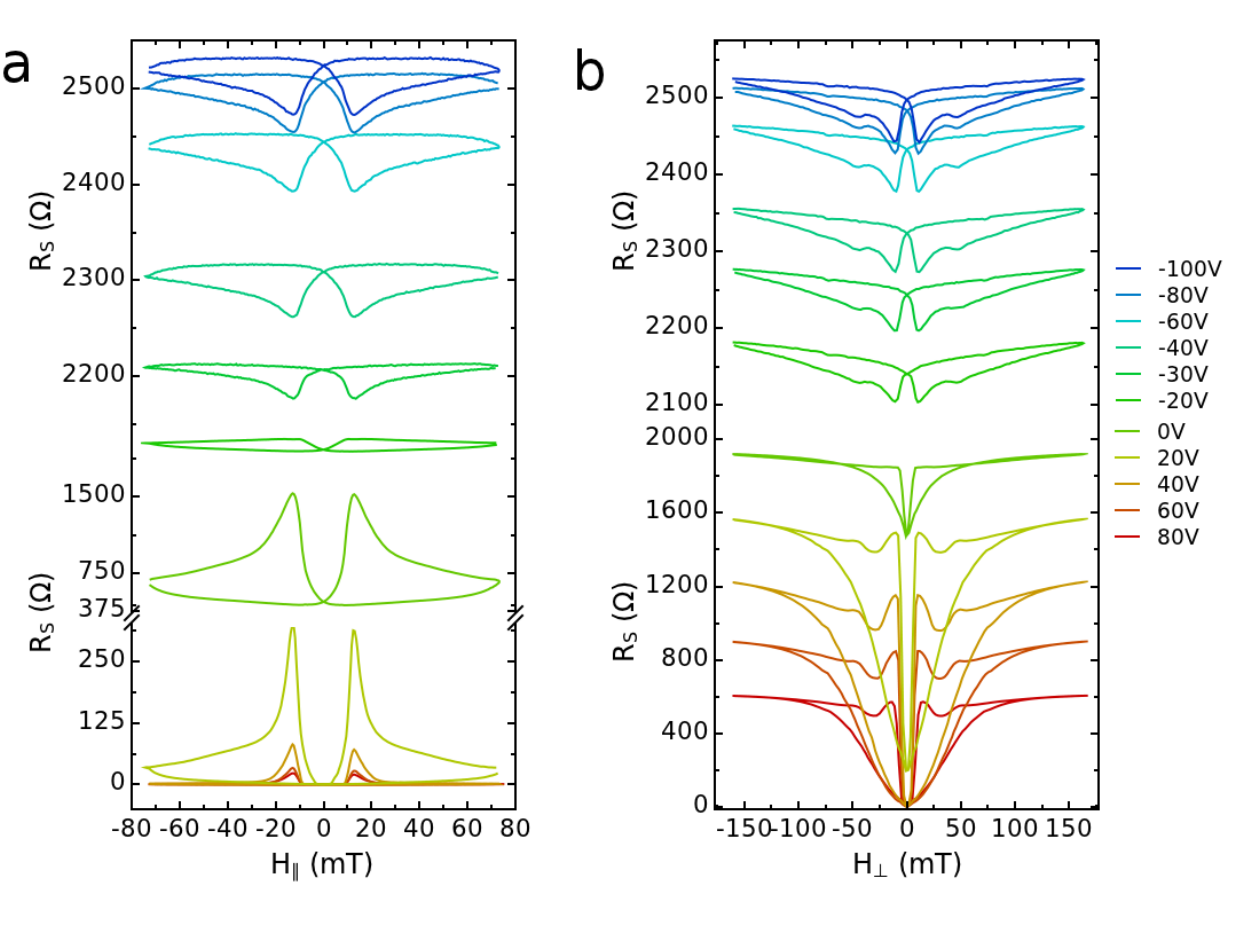}

%\vspace{-1cm}
\caption{\textbf{SIT in the magnetoresistance.} \textbf{a,} Parallel 
field
MR for the fastest sweep rate (300 $\mu$T/s) as a function of gate voltage, tuning the system from a superconducting to an insulating 
regime. The top and bottom panels are for insulating and 
superconducting regimes respectively. Charge-vortex duality manifests itself as the conversion of the \textit{peak} in the superconducting 
regime to a \textit{dip} in the insulating
regime. The peak and dip occur at the external 
field value of $H_{\parallel} \sim \pm$ 12 mT. (There is an axis break on the y-axis in the bottom panel.) 
The maximum change 
in the resistance occurs for $V_g =$ 0 V, where the superconductivity is very weak. \textbf{b,} Similar behaviour in perpendicular field. 
Note that the fastest sweep rate in this case is 600 $\mu$T/s. Additional
structure in the perpendicular field at $H_{\perp} \sim \pm$ 48 mT is due to the more complex magnetization dynamics in the perpendicular field.} 
\end{center}
\label{fig3}
\end{figure}  

\begin{figure}[h]
\begin{center}
\includegraphics[width=0.9\textwidth]{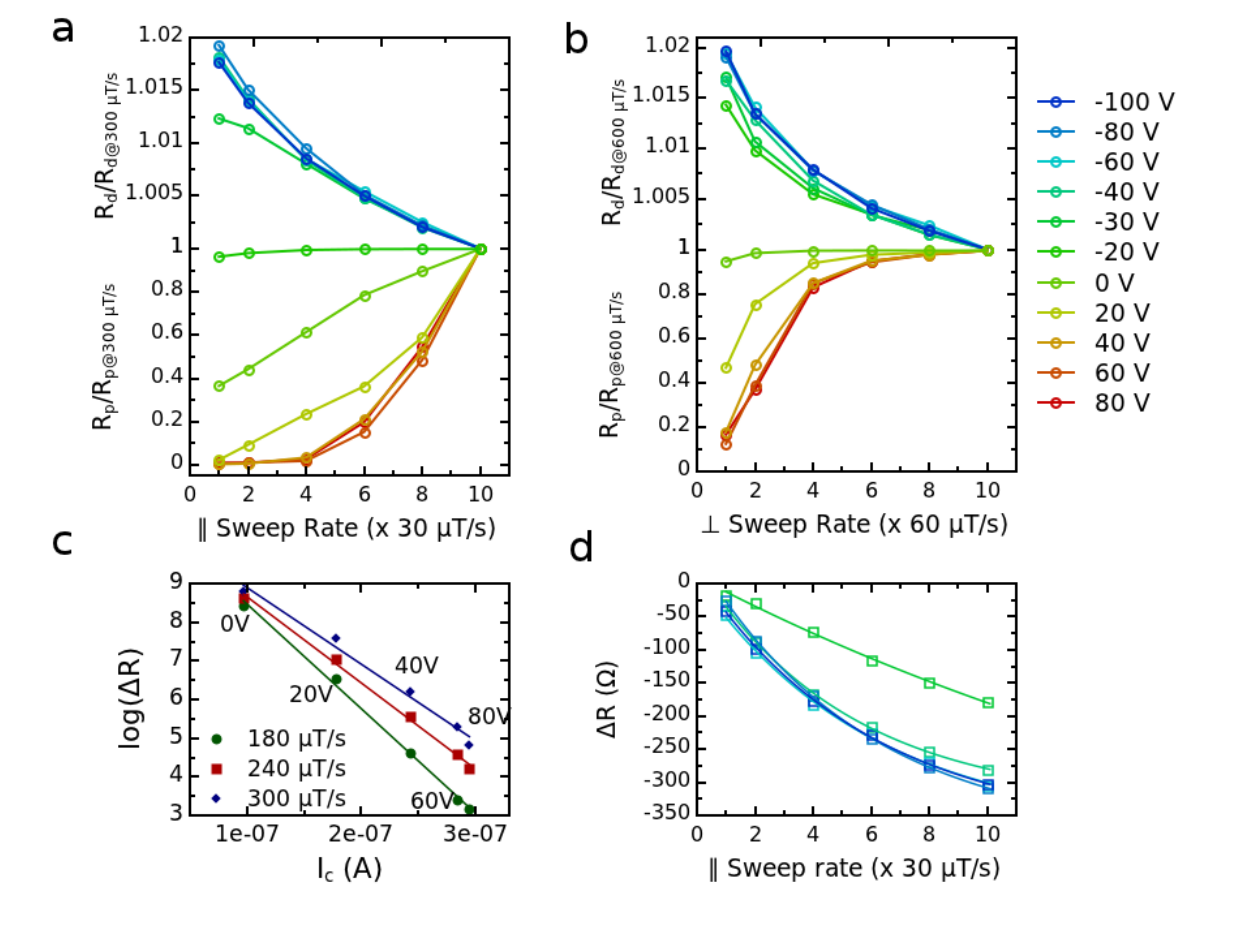}

%\vspace{-1cm}
\caption{\textbf{Sweep rate dependence.} \textbf{a,} The evolution of the resistance of the peaks and dips (normalized to the peak/dip value at the 
highest sweep rate) at $H_\parallel \sim$ 12 mT with 
external field sweep rate as a function of the gate voltage for parallel field. The top and bottom panels are for insulating and 
superconducting regimes respectively. There is no saturation of 
the resistance of the peaks in the 
superconducting regime for the parallel field as the only contribution to pair destruction is from the field due to 
the domain wall. \textbf{b,} Similar plot for perpendicular field. The resistance of the peaks quickly saturates in the superconducting 
regime due to the pair breaking effect of the external field, unlike in the parallel field case. However, the resistance of the dips on the 
insulating side does not saturate. \textbf{c,} A plot of $log(R_{peak} - R_{no peak})$ vs the critical current, $I_c$, for the three fastest sweep 
rates in external field. $R_{no peak}$ is defined as the resistance at the same field value as $R_{peak}$ but on the field sweep curve in the 
opposite direction. The three lines are the fits to the data at the three different sweep rates. \textbf{d,} $R_{dip} - R_{no dip}$ as a function of the parallel 
field sweep rate. The data for each gate voltage, $V_g$ = -30, -40, -60, -80, and -100 V are fitted to an exponentially decaying function. }
\end{center}
\label{fig4}
\end{figure}

\begin{description}
  \item \textbf{Acknowledgements} \\
  We would like to thank P. A. Lee, A. M. Goldman and J. B. Ketterson for providing useful comments. Work at Northwestern was supported by a grant from the DOE
  Office of Basic Energy Sciences under grant No. DE-FG02-06ER46346.  Work at the University of Wisconsin was supported by funding from the DOE Office of Basic Energy Sciences under award number DE-FG02-06ER46327 and the National Science Foundation under Grant No. DMR-0906443.   
  \item \textbf{Correspondence} \\ 
  Correspondence should be addressed to V.C. (v-chandrasekhar@northwestern.edu). 
\end{description}

\end{document}

\title{Supplementary Information for Evidence for charge-vortex duality at the LaAlO$_3$/SrTiO$_3$ interface}

%\author{M. M. Mehta,$^1$ D. A. Dikin,$^1$  C. W. Bark,$^2$ S. Ryu,$^2$ C. M. Folkman,$^2$  C. B. Eom$^2$ and V. Chandrasekhar$^1$ }

%\footnotetext[1]{Department of Physics and Astronomy, Northwestern
%University, Evanston, IL 60208, USA}

%\footnotetext[2]{Department of Materials Science and Engineering
%University of Wisconsin-Madison, Madison, WI 53706, USA}

\baselineskip24pt

\maketitle 

\section{Experimental Techniques}

The samples in this work had 10 unit cells (uc) of LaAlO$_3$ (LAO) grown by pulsed laser deposition on TiO$_2$ terminated (001) SrTiO$_3$ (STO) 
single crystal substrates \cite{park,bark}.  The electrical measurements were performed on a Hall bar defined by photolithography and etched using 
argon ion milling.  Details of the film preparation, characterization and sample fabrication have been discussed in detail in prior publications 
\cite{park,bark,dikin}.  
The samples were measured in an Oxford dilution refrigerator with a base temperature of 15 mK.  This refrigerator was equipped with a 2-axis 
magnet so that a magnetic field could be applied both perpendicular and parallel to the LAO/STO interface.  A gate voltage $V_g$ was applied to the 
back of the 500 $\mu$m thick STO substrate. 

Transport measurements were made by a standard ac lock-in detection method. The samples were current biased with an excitation current of 
$I_{ac}$ = 10 nA at a frequency of 11.3 Hz. For $V_g =$ 20 V and higher, this excitation current is much less than the 
critical current, $I_c$ ($I_{ac} < 0.1I_c$). The voltage signals were first amplified 
using a low noise homemade instrumentation amplifier and then measured using a PAR 124 analog lock-in amplifier or 
an EG\&G 7260 digital lock-in amplifier.

\section{Determination of film thickness and parallel critical field} 

From the continuous mapping of $T_c$ vs $H_{\perp}$ one can estimate the superconducting coherence length, $\xi$, as discussed in 
Ref. \ref{dikin}.
We obtained a value for the coherence length of $\xi \sim$ 70 nm at $V_g =$ 80 V. In order to determine the thickness of the 
superconducting layer, we performed a similar measurement of the dependence of $T_c$ on $H_{\parallel}$. The system was biased at the mid-point
of the resistive transition ($R_S = $ 288 $\Omega$ @ $V_g =$ 80 V), under a proportional-integral-differential (PID) feedback circuit. 
The output of the PID was used to drive the mixing chamber heater 
while the field was ramped. This ensured that we always stayed at $T_c$, defined as the mid-point of the resistive transition. A plot of 
$T_c$ vs $H_{\parallel}$ measured in this way is shown in Fig S1. 
As expected for a 2D superconductor, the dependence of $T_c$ on a parallel applied field 
is quadratic except at low fields where hysteresis due to the ferromagnet suppresses $T_c$. From a fit to this measurement at high fields 
we extracted the parallel 
critical field, $H_{c\parallel} \sim$ 1.2 T. For a thin film superconductor, $H_{c\parallel}$ and the film thickness are related by the formula \cite{tinkham}

\begin{equation}
 H_{c\parallel} = \frac{\sqrt{3}\Phi_{\circ}}{\pi d\xi}
\label{eqnsuppl1}
\end{equation}
where $\Phi_{\circ} = h/2e$ is the superconducting flux quantum. 

From a knowledge of $\xi$ and $H_{c\parallel}$, we determined the thickness of the superconducting film 
to be $d =$ 13.6 nm, which is less than $\xi$.

\begin{figure}[h]
\begin{center}
\includegraphics[width=10cm]{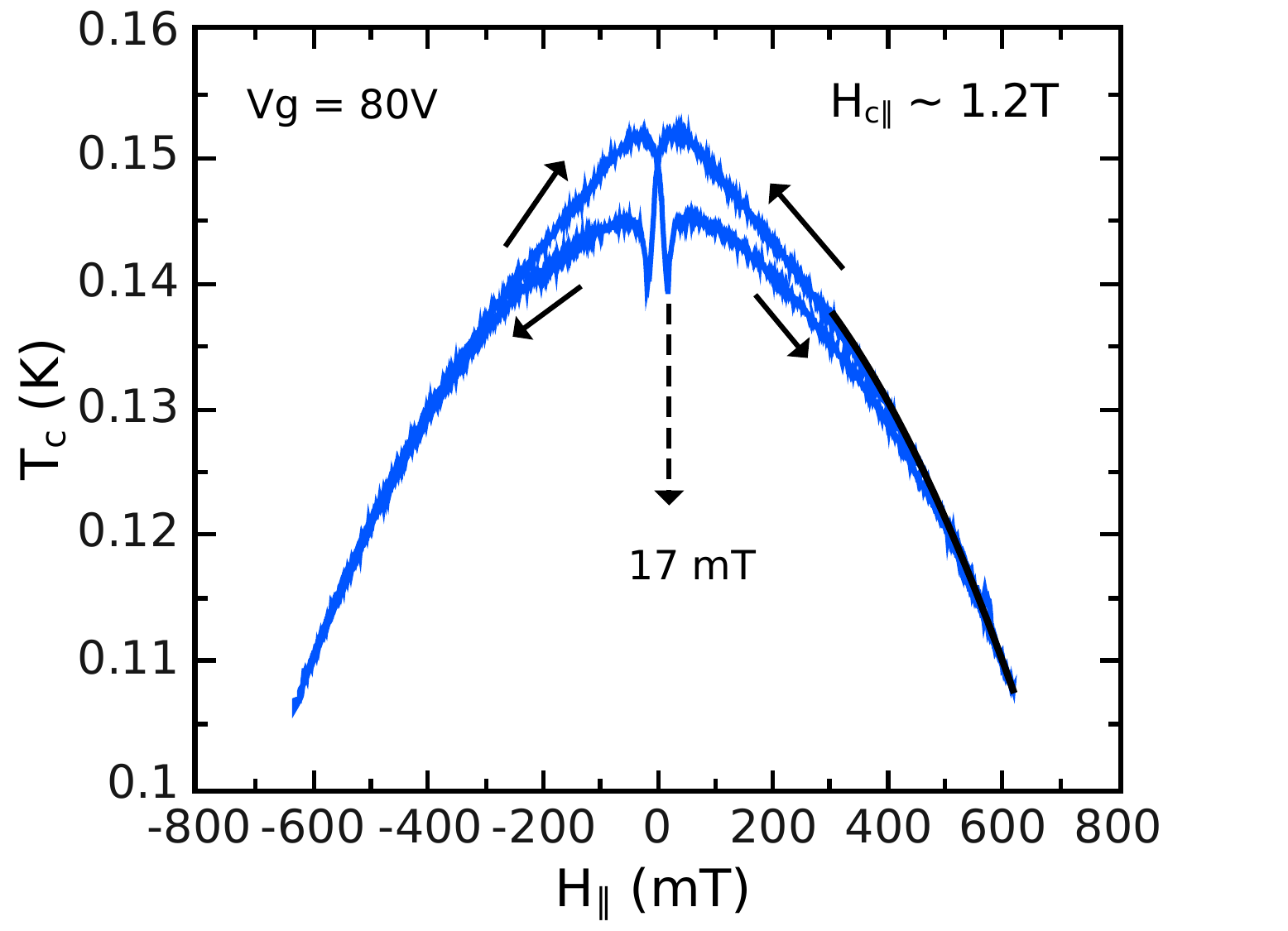}
\caption*{\textbf{Figure S1: Parallel critical field} $T_c$ vs $H_{\parallel}$ for $V_g =$ 80 V. A parabolic fit (black curve) at high fields gives $H_{c\parallel} \sim$ 1.2 T.
The thickness of the superconducting layer estimated from this value is $d =$ 13.6 nm.}
\end{center}
\label{figS1}
\end{figure}  

\section{Magnetoresistance in crossed parallel and perpendicular fields}

Since our system shows hysteresis due to the magnetization dynamics of the ferromagnet, it becomes inherently more complex to study the effect 
of an externally applied field to the system. In order to eliminate the hysteresis we applied a persistent parallel field, $H_\parallel$, 
which saturated the magnetization of the ferromagnet in one direction, and then performed perpendicular field magnetoresistance measurements. 
In Fig. S2 we show data in which perpendicular magnetoresistance is measured for various values of $H_\parallel$ for $V_g =$ 100 V at $T = $ 50 mK. 
It can be seen that for $H_{\parallel} =$ 20 mT and higher, the sharp resistance peak in the magnetoresistance is absent and the magnetization 
of the ferromagnet is completely saturated. For $H_{\parallel} =$ 0 mT, the sharp peak is recovered. Note that the minimum of resistance and 
the field at which this minimum occurs both increase as the parallel field is increased. The additional pair breaking caused due to a constant 
parallel field gives rise to the small increase in resistance. The shifting of the zero of the magnetic field is due to a small misalignment 
of plane of the sample with respect to the axis of our split coil magnet. From the shift we calculate the misalignment to be 1.4$^{\circ}$. 

\begin{figure}[h]
\begin{center}
\includegraphics[width=10cm]{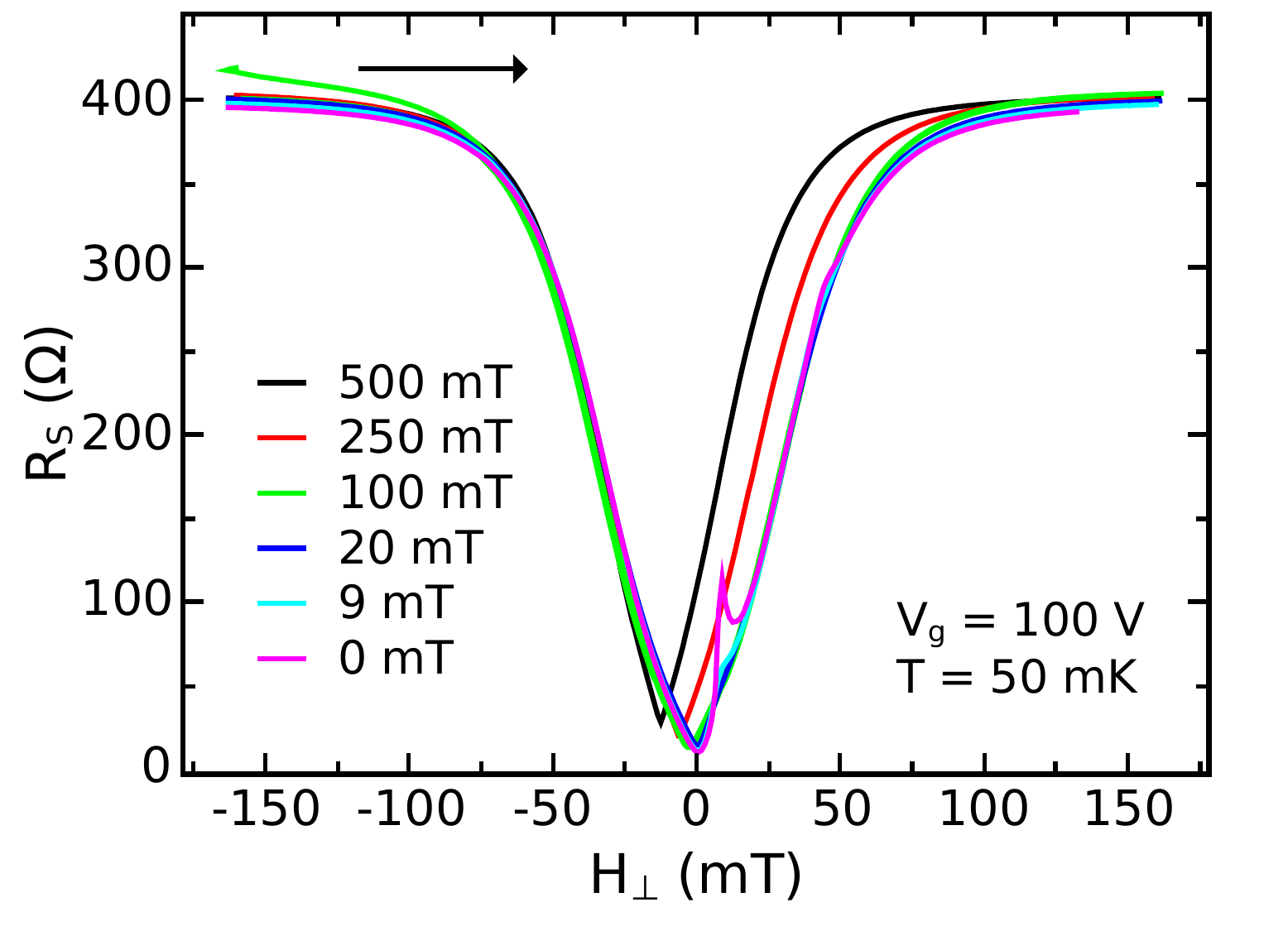}
\caption*{\textbf{Figure S2: Crossed parallel and perpendicular field} $R_S$ vs $H_{\perp}$ for different values of parallel fields at $V_g =$ 100 V. For clarity field sweeps
in only one direction are shown, with the arrow indicating the direction of field sweep.}
\end{center}
\label{figS2} 
\end{figure}   

In the insulating phase (for $V_g$ = -40 V, say) the generation of flux due to domain wall motion 
is responsible for a rise in conductance. As is discussed in the main text and in detail below, a changing vector potential, $\partial A/\partial t$, generates 
an electric field, $\mathcal{E}_H$, which causes a potential, $V$, to develop between two adjacent islands. When this potential energy $2eV$ 
exceeds the charging energy of the islands, $E_c$, one gets conduction between the islands. In order to see that the dips in resistance that 
we see on the insulating side are indeed due to such a mechanism, we show $R$ vs $H_{\perp}$ traces for $V_g =$ -40 V, for $H_{\parallel} =$ 0 
and 500 mT in Fig S3. Both the traces in Fig S3 are at the same sweep rate. By applying 
a parallel field of 500 mT, we have eliminated the effect of the magnetization dynamics in the ferromagnet. The sharp dips in resistance vanish 
completely, and only a small positive magnetoresistance due to weak localization remains. The shift in zero of the $H_\parallel =$ 500 mT curve 
is due to the small misalignment of the sample as discussed above.

\begin{figure}[h]
\begin{center}
\includegraphics[width=10cm]{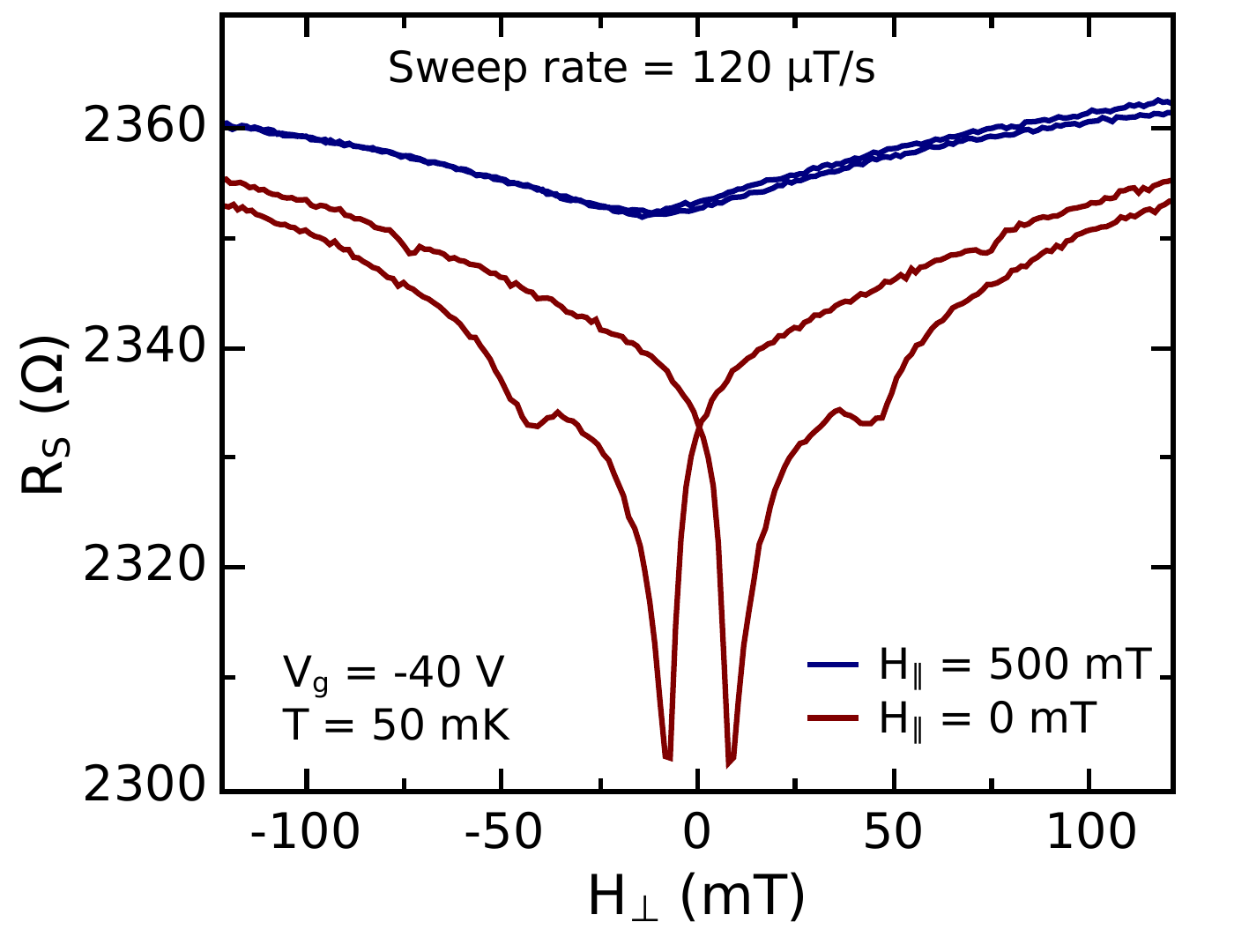}
\caption*{\textbf{Figure S3: Charge-vortex duality}  $R_S$ vs $H_{\perp}$ for $H_{\parallel} =$ 0 and 500 mT at $V_g =$ -40 V. For both curves, the field is swept
at the same rate. For $H_{\parallel} =$ 500 mT, characteristic features due to the ferromagnet are absent.}
\end{center}
\label{figS3}
\end{figure} 

\section{Insulating state I-V characteristics}

The I-V characteristics of the system were discussed in Ref. \ref{dikin}. The I-V curves in the insulating state were almost linear with no 
clear evidence of a Coulomb gap, indicating a weak insulating state. However, if $dV/dI$ ($R_S$) is plotted as a function of an applied bias 
current $I_{dc}$, a peak in $dV/dI$ is observed at low bias. This is shown in Fig S4 for $V_g =$ -30, -45, -60 and -100 V at $T =$ 15 mK. 
For $V_g =$ -30 and -45 V, a dip in $dV/dI$ at zero bias signifies the presence of some superconductivity, however no evidence for that is 
seen in the temperature dependence of the resistance since $dR/dT <$ 0 at all temperatures in the temperature range of the current experiments. The dip in $dV/dI$ probably 
arises due to localized superconducting islands weakly phase coupled by the Josephson effect. For $V_g =$ -60 and -100 V, 
the system is in the insulating regime with a peak in $dV/dI$ at zero bias. 

\begin{figure}[h]
\begin{center}
\includegraphics[width=10cm]{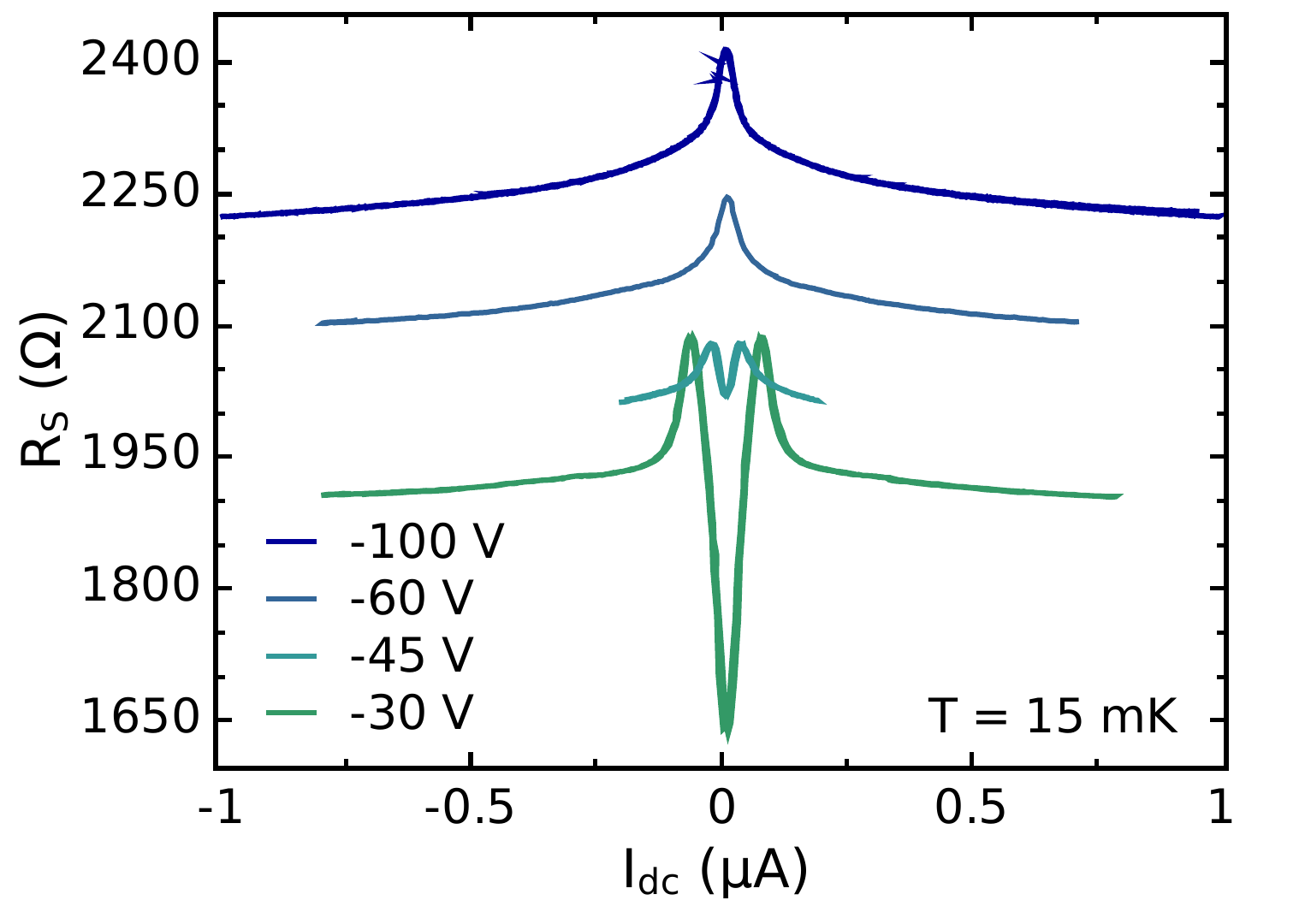}
\caption*{\textbf{Figure S4: Insulating state I-V characteristics}  $R_S$ vs $I_{DC}$ for different gate voltages at $T =$ 15 mK. }
\end{center}
\label{figS4}
\end{figure}   

\section{Weak localization effects in the insulating state}

It could be argued that the rise in resistance with decreasing temperature in the insulating regime is due to weak localization effects in the 
electron gas. For a two-dimensional metal, the quantum interference contribution to the magnetoresistance including spin-orbit 
scattering can be written as\cite{santhanam} 
\begin{equation}
\frac{\Delta R}{R}^{loc}(H) = \frac{R_S}{2 \pi^2 \hbar/e^2} \left[\left(-\frac{3}{2}f_0(H_2) + \frac{1}{2} f_0(H_\phi)  \right)\right],
\label{eqnsuppl2}
\end{equation} 
where 
\begin{equation}
f_0(H_i) = \psi(1/2 + H_i/H)  - \ln(H_0/H),
\label{eqnsuppl3}
\end{equation}
$\psi$ is the digamma function,  $H_0 = \Phi_0/8 \pi \ell^2$, $H_2 = H_\phi + (4/3) H_{so}$, $H_\phi = \Phi_0/8 \pi L_\phi^2$ 
and $H_{so}=\Phi_0/8 \pi L_{so}^2$.  
$\ell$ is the elastic mean free path, $L_\phi$ is the electron phase coherence length and $L_{so}$ is the spin-orbit scattering length.

Experimentally, the measured quantity is $\delta R (H)/ R$ = $(\Delta R/R)(H) - (\Delta R /R)(0)$.  Taking the appropriate limits in 
Eqn. \ref{eqnsuppl1} above, one obtains:
\begin{equation}
\frac{\delta R}{R}^{loc}(H) = \frac{R_S}{2 \pi^2 \hbar/e^2} \left[-\frac{3}{2} f_1(H_2) +\frac{1}{2} f_1(H_\phi)  \right]
\label{eqnsuppl4}
\end{equation}
where
\begin{equation}
f_1(H_i) = \psi(1/2 + H_i/H)  - \ln(H_i/H).
\label{eqnsuppl5}
\end{equation}
It should be noted that Eqn. \ref{eqnsuppl3} does not depend on $H_0$, and consequently the elastic mean free path does not enter into the analysis.

Figure S5 shows the $\delta R (H)/ R$ for $V_g =$ -100 V at $T =$ 50 and 400 mK. The curves at both the temperatures are reasonably well fit  by Eq. \ref{eqnsuppl4}. From the fitted curves one obtains at $T =$ 50 mK, $L_\phi =$ 93 nm and $L_{so} =$ 53 nm, and at $T =$ 400 mK, 
$L_\phi =$ 60 nm and $L_{so} =$ 42 nm. Using these values the WL contribution to the rise in resistance with decreasing temperature,
$\delta R(T,H \rightarrow 0)$, is estimated to be $\delta R(T,H \rightarrow 0) \sim$ 200 $\Omega$ as $T$ is reduced from 400 mK to 50 mK. However, the 
observed rise in resistance at $V_g =$ -100 V between these two temperatures is $\sim$ 800 $\Omega$. Hence the increase in resistance cannot be 
solely accounted for by WL.

\begin{figure}[h]
\begin{center}
\includegraphics[width=10cm]{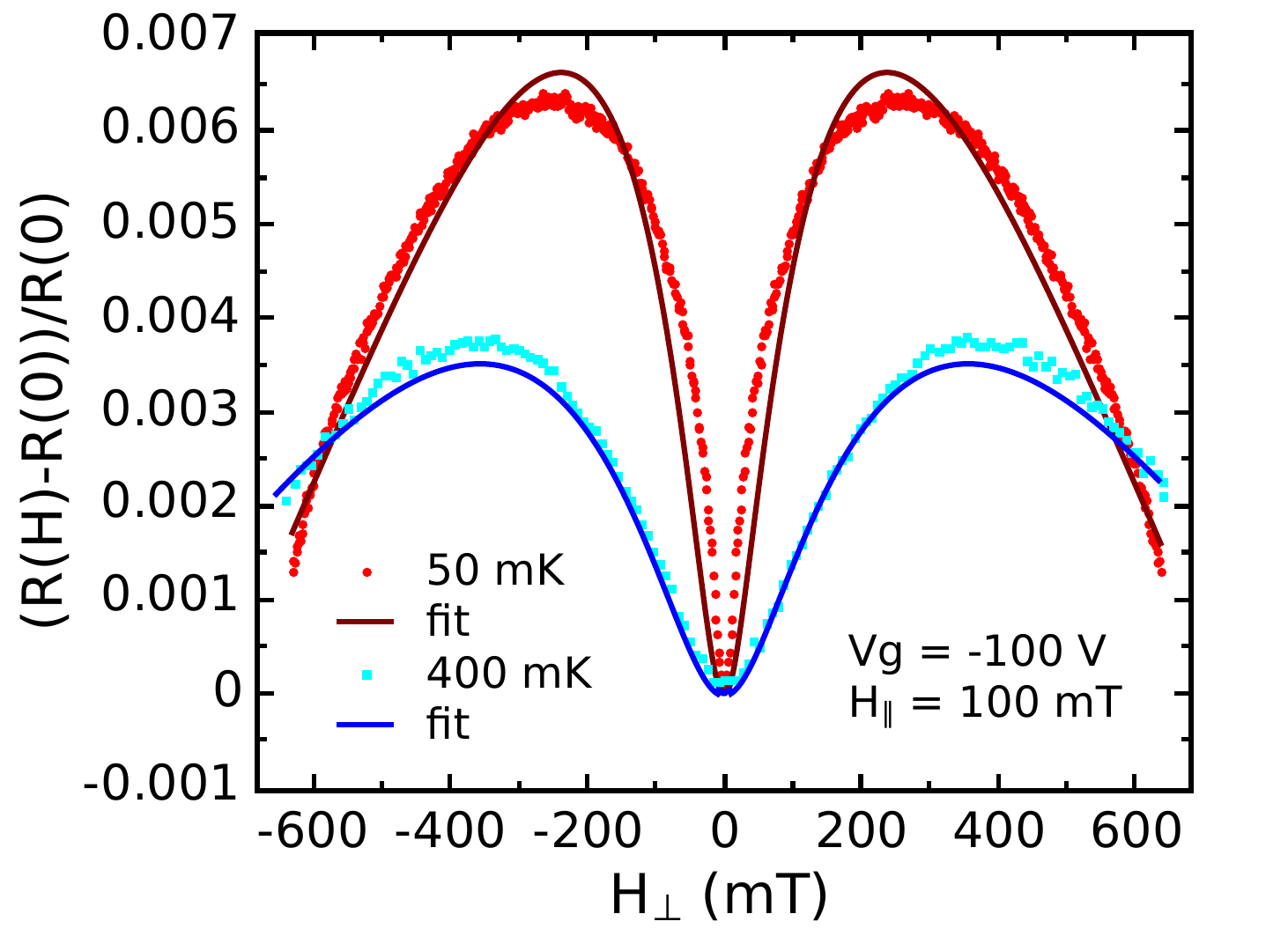}
\caption*{\textbf{Figure S5: Weak localization analysis}  $R(H)-R(0)/R(0)$ vs $H_\perp$ for $V_g =$ -100 V at $T =$ 50 (red circles) and 400 (cyan squares) mK 
in a constant parallel field, $H_\parallel =$ 100 mT. The solid lines are the fits to the two curves using Eqn. \ref{eqnsuppl4}. }
\end{center}
\label{figS5}
\end{figure}   

The contribution to resistance due to electron-electron ($e-e$) interactions for a 2D metal is logarithmic in temperature and is given by \cite{santhanam}

\begin{equation}
\Delta R_S^{ee}(T) = -\frac{\alpha R_S ^{2}}{h/e^2}\ln\left(\frac{T}{T_{ref}} \right),
\label{eqnsuppl6}
\end{equation} 
where $T_{ref}$ is some high temperature value for reference and $\alpha$ is a measure of the strength of the $e-e$ interactions; typically 
$\alpha \sim \mathcal{O}$(1). 
If we take $T_{ref} =$ 400 mK and calculate the $e-e$ interaction contribution to the
resistance at $T =$ 50 mK, we get $\Delta R^{ee}(T = $50 mK$) = \alpha\times$505 $\Omega$. Since the WL contribution $\sim$ 200 $\Omega$, and the total
change in resistance $\sim$ 800 $\Omega$, we can conclude that in our case $\alpha \gtrsim \mathcal{O}$(1). From the above analysis, it can be 
concluded that the major contribution to the rise in resistance with decreasing temperature in the insulating regime is due to $e-e$ interactions. 

\section{Thermal activation model for the resistance peaks/dips}
The model that we have for our sample is a random network of superconducting islands that are coupled with a characteristic Josephson energy $E_J$.  
The islands are small enough that there is a Coulomb penalty for transferring Cooper pairs from one island to another, characterized by the charging 
energy $E_c$.  Of course, for a random network of islands, there will be a distribution of Josephson couplings and charging energies, so $E_J$ and $E_c$ 
represent averages over the sample. 

\subsection{Superconducting regime}
In this picture, on the superconducting side of the transition in the absence of motion of magnetic vortices, there is strong Josephson coupling between 
the islands ($E_J>>E_c$), and the sample is in the zero resistance state.  Any magnetic vortices present are localized in the interstitial regions 
between the islands.  Motion of the magnetic vortices, which in our case is induced by the magnetization dynamics in the ferromagnet, will give rise 
to a finite resistance as the vortices cross the junctions between the superconducting islands.  In order to understand the mechanism of the generation 
of this resistance, we use the phase slip model due to Langer and Ambegaokar (LA) \cite{langer}.   There are corrections to this model, for example 
by McCumber and Halperin \cite{mccumber},  but the basic concepts are all that is essential for our analysis.

To break down the problem, we consider two adjacent superconducting islands that are Josephson coupled to each other at two points, so that the 
interstitial region between them can enclose a magnetic vortex :  the sample consists of a network of such interconnected islands.  
(More generally, one probably has multiple interconnected islands enclosing vortices, but this does not change the physical picture.)  
These two islands can be thought of as a dc SQUID, whose energy profile is given by
\begin{equation}
 U(\phi) = -E_J\cos \phi 
 \label{eqnsuppl7}  
\end{equation}
where $\phi$ is the phase difference across the SQUID and $E_J$ is the Josephson energy.  $E_J$ is proportional to the critical current $I_c$ of the SQUID. 
This defines a periodic potential where the minima of the potential correspond to integral values of the phase $\phi = 2n \pi$.  At 
finite temperature, the system can be thermally activated from one potential minimum to an adjacent minimum which differs by a 
value $\delta \phi = \pm 2 \pi$ over the energy barrier represented by $E_J$.  Each such phase slip event will give rise to a voltage pulse 
according to the Josephson relation
\begin{equation}
 V = \frac{\hbar}{2e}\frac{d\phi}{dt}
 \label{eqnsuppl8}  
\end{equation} 
where the phase change is $\pm 2 \pi$.  In the absence of an external current, phase jumps in either direction are equally likely, hence the average 
voltage measured across the junction is still zero, although such phase slip events may give rise to voltage noise.
 
In the presence of an external measuring current $I$, the SQUID's energy profile is modified to the ``tilted washboard'' potential 
\begin{equation}
 U(\phi) = -E_J\cos \phi - \frac{\hbar}{2e}I\phi .
 \label{eqnsuppl9}  
\end{equation}
In this case, phase slip events corresponding to the system traveling ``down'' the washboard tilt are slightly more likely than those in the other 
direction, leading to a finite average voltage, and hence a finite resistance.  However, the probability of such events is still quite small at 
low temperatures if $I<<I_c$.

LA's original paper focused on phase slips in a single weak link between two superconductors.  They derived the rate at which phase slips 
occur in each direction
\begin{equation}
 \eta_{\pm} = \Omega e^{-\Delta F^{\pm}/k_BT} 
 \label{eqnsuppl10}  
\end{equation}
where the attempt
 frequency $\Omega$ depends on microscopic parameters as well as external parameters of the circuit.  Here $\Delta F^\pm$ is the free energy barrier 
for phase jumps in the 
the two directions \cite{halperin}
\begin{equation}
 \Delta F^{\pm} \approx 2E_J \pm \frac{\pi \hbar I}{2e} .
 \label{eqnsuppl11}  
\end{equation}
In the LA model, the application of a voltage $V$ drives the generation of phase slips, and in steady state, the average voltage due to the 
generation of phase slips is equal to the applied voltage, giving the relation 
\begin{equation}
  V = (2\pi \hbar /e)\Omega e^{-2E_J/k_BT}\sinh (\pi \hbar I/2ek_BT) 
 \label{eqnsuppl12}  
\end{equation}
In the low current limit ($I \ll ek_BT/\hbar$) the corresponding resistance is
\begin{equation}
 R =  V /I = (\pi \hbar^2 /e^2k_BT)\Omega e^{-2E_J/k_BT} .
 \label{eqnsuppl13}  
\end{equation}

In our case, a phase slip is generated whenever a magnetic vortex line crosses a weak link.  The rate at which these phase slips are generated 
would then be proportional to the time dependence of the field generated by the magnetization of the ferromagnet.  At the field sweep rates in 
our experiment, the magnetization change is essentially quasi-static, hence this time dependence is proportional to the external magnetic field 
sweep rate $\dot{B}$, resulting in an additional factor in Eqn(\ref{eqnsuppl13}) proportional to $\dot{B}$.   The change in resistance due to the 
motion of vortices for the entire sample is an average of terms such as Eqn (\ref{eqnsuppl13}) with a distribution of $E_J$'s, but will be 
proportional to the magnetic field sweep rate $\dot{B}$.  The measured value of the critical current $I_c$ of the sample will be determined by a 
parallel combination of random paths through the weak links connecting the superconducting islands.  Nevertheless, it is reasonable to assume that 
this measured critical current $I_c$ is proportional to the mean Josephson energy $E_J$ for the superconducting network.  Thus, we can write the 
change in resistance due to the motion of magnetic vortices in the form
\begin{equation}
  \Delta R = A \dot{B} e^{-\alpha I_c/k_B T}
  \label{eqnsuppl14}
\end{equation}
where $I_c$ is the measured critical current, and $A$ and $\alpha$ are numerical constants at a fixed temperature.  Thus, if one 
plots $\ln(\Delta R)$ as a function of $I_c$ at a fixed temperature, one should obtain a straight line.  $I_c$ in this system can be varied by 
changing the gate voltage $V_g$.  Figure 4(c) in the main text shows this plot.  As can be seen, the expected exponential dependence is indeed 
observed for a number of different magnetic field sweep rates.  The exponential factor $\alpha$ for the different sweep rates (which should be the same) 
matches to within about 20 \%.

Equation (\ref{eqnsuppl14}) also predicts that the resistance change should be proportional to the magnetic field sweep rate $\dot{B}$.  If one 
looks at the parallel field data for  $V_g$= 0 and 20 V shown in Fig. 4(a) of the main text, one can see that the dependence on sweep rate is indeed 
approximately linear.  However, for $V_g$=  40,  60, 80 V, which are deeper in the superconducting regime, the dependence is clearly not linear.  
In fact, deep in the superconducting regime, one might expect very little change in resistance, as the exponential factor involving $I_c$ in 
Eqn (\ref{eqnsuppl14}) would suppress any resistance change.  Thus the data at the three slowest sweep rates is not surprising.  However, at higher 
sweep rates, the peak amplitude for these gate voltage values does increase.  We do not know the reason for this, although one possibility is that 
at higher sweep rates, the effective Josephson coupling between islands is modified.  We note that the data for the perpendicular field 
shown in Fig. 4(b) of the main text do not show a linear behavior, but in this case, the situation is complicated by the fact that there is also 
a orbital contribution to the peak height due to the external perpendicular magnetic field.

\subsection{Insulating regime}  
As discussed in the main text, on the insulating side of the transition, charge transport between islands is suppressed due to the charging 
energy $E_c$.  If one considers thermal activation, the rate at which charge can be transferred is proportional to $e^{-E_c/k_B T}$, and as with 
the superconducting regime above, charge transfers in both directions are equally likely, and there is no net average current.  In the presence of a 
finite voltage difference $\Delta V$ between the two islands, the rate is modified to
$e^{-(E_c \pm 2e\Delta V)/k_B T}$, and consequently favors charge transfer from the  island at higher potential to the island at lower potential, 
resulting in a net average current between the islands.  In our model, the potential $\Delta V$ results from the electric field that is generated by the 
moving magnetic field generated by the magnetization dynamics in the ferromagnet through Faraday's law.  The time dependence of the changing 
magnetization, as we have pointed out above, is proportional to the sweep rate of the external magnetic field $\dot{B}$, and 
hence $\Delta V = \gamma \dot{B}$, where $\gamma$ is a constant. 
Thus, the resistance of the dip should have a dependence on the magnetic sweep rate of the form
\begin{equation}
\Delta R = - e^{\gamma \dot{B}/k_B T}.
\label{eqnsuppl15}
\end{equation}
We note that this is a characteristically different dependence than that which is predicted and found in the superconducting regime.
 
Figure 4(c) in the main text shows a plot of the resistance change at the dip as a function of the sweep rate for different gate voltages on the 
insulating side:  the lines are fits to the exponential form Eqn. (\ref{eqnsuppl15}).   As can be seen, the fits are quite good.  
For $V_g$= -100, -80, -60 V, and -40 V the value of $\gamma$ is almost the same.  For $V_g$= -30 V, it is quite different.  The time dependent magnetic 
field generates an electric field; the resulting voltage $\Delta V$ is proportional to the the electric field, but also to the spacing $d$ between 
the islands.  Hence $\gamma$ should also depend on $d$.  As $V_g$ is changed to bias the system closer to the transition from the insulating side, 
it is reasonable to expect that the effective spacing between islands decreases, resulting in a decrease in $\gamma$, as is observed in Fig. 4(c) 
of the main text. 

\section{Alternative scenarios for the dip in the MR in the insulating regime}

As we cannot drive the system deep into the insulating regime by changing $V_g$, one might argue that the model of isolated superconducting islands is 
not applicable, and one has instead a weakly localized metal.  However, it is hard to explain the origin of the sweep-rate dependent dips in the MR 
for a metal.  For example, if one had a metal, one possible scenario is that the moving magnetic field associated with the domain wall leads to 
momentary eddy current heating of the electron gas, resulting in a decrease in resistance as $dR_S/dT <$ 0 in the insulating regime in our temperature 
range (Ref. \ref{dikin}). However, as the resistance of the film is large, the eddy current heating is expected to be small.  In addition, a detailed 
comparison of the magnitude of the dips in the MR in comparison with the change in resistance as a function of temperature argues against this mechanism. 
Assuming that the moving magnetic field results in eddy current heating of the electron gas, one can estimate the increase in electron temperature by 
mapping the magnitude of the resistance change at a dip at the fastest sweep rate in parallel field, $R_{dip} - R(0)$ onto the $R$ vs $T$ curve\cite{dikin}.  
Doing this, one obtains an estimated increase in electron temperature of $\Delta T \sim$ 65 mK for $V_g =$ -30 V, as shown in Fig. S6. However, similar 
heating effects should also be observed for gate voltages at which the sample is resistive, but shows a decrease in resistance , for e.g., $V_g =$ -20 V, 
which shows a peak in the MR (Fig. 3a of the main text).   Indeed, eddy current heating in this $V_g$ range should be larger, as the resistance is lower.  
For $V_g =$ -20 V, $R_{peak} - R(0) =$ 620 $\Omega$; however, the $R$ vs $T$ curve for $V_g =$ -20 V does not show a corresponding increase in 
resistance (Fig. S6). Clearly for $V_g =$ -20 V neither the sign nor the magnitude of $\Delta R$ is consistent with the eddy current heating scenario.  
We have attempted to come up with other potential mechanisms to explain the MR dips on the insulating regime, but there appears to be no other explanation 
that agrees with the experimental data other than the one that is discussed in the main text.

\begin{figure}[h]
\begin{center}
\includegraphics[width=10cm]{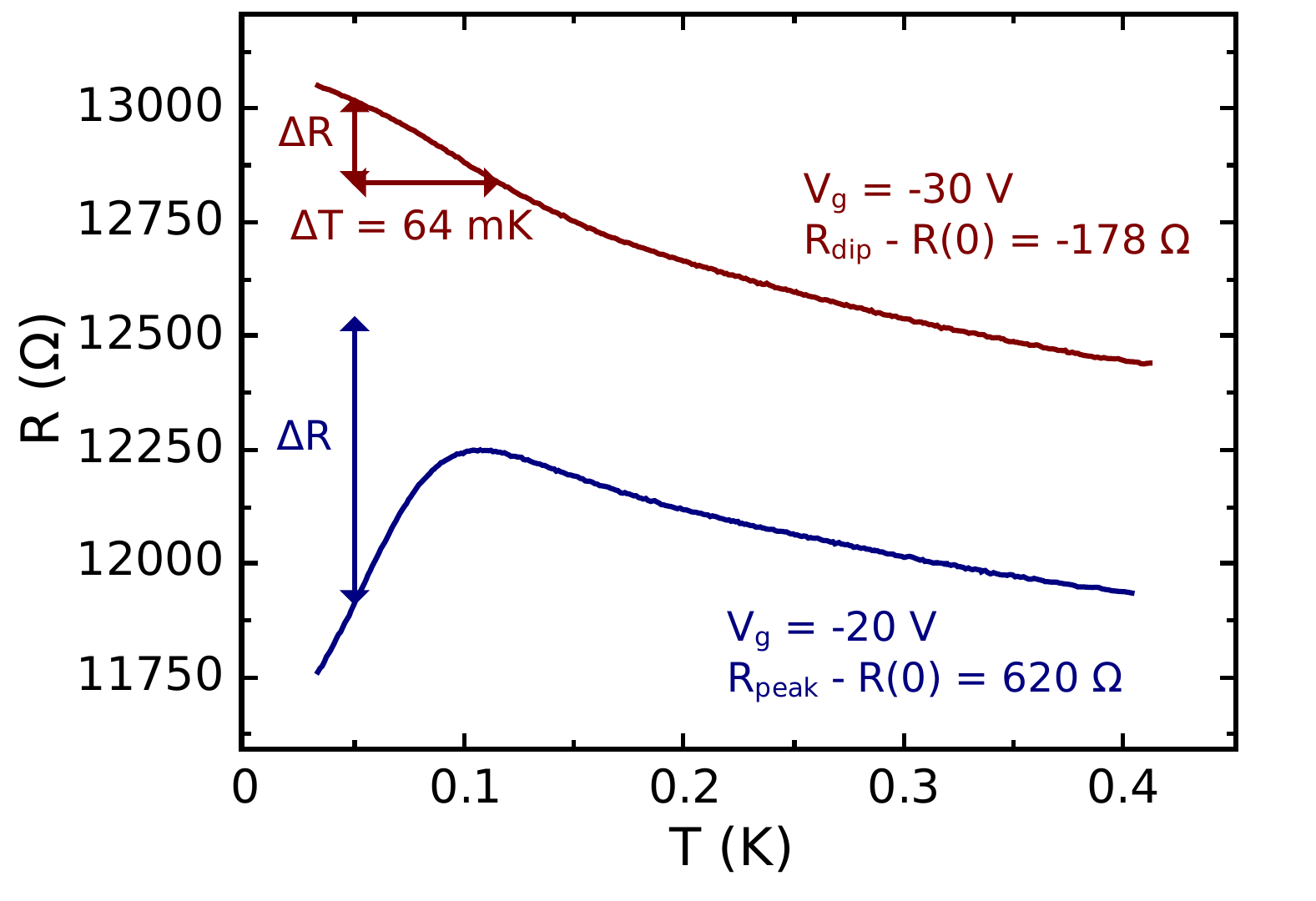}
\caption*{\textbf{Figure S6: Eddy current heating}  $R$ vs $T$ for $V_g =$ -30 and -20 V. The vertical arrows indicate the size of $\Delta R$ ($R_{peak/dip} - R(0)$)
corresponding to the change in resistance at the peak or dip in the MR; the horizontal arrow shows the expected change in temperature if the dip in the MR is caused due to the eddy current heating.   The explanation clearly does not work for the lower curve, as the change in resistance at the peak is larger than the change in resistance due to temperature.}
\end{center}
\label{figS6}
\end{figure}